\documentclass{aa}
\usepackage{graphicx}

\def\jh{\mbox{$\rm (J-H)$}}
\def\ebv{\mbox{$\rm E(B-V)$}}
\def\ejh{\mbox{$\rm E(J-H)$}}
\def\rc{\mbox{$\rm R_{core}$}}
\def\rl{\mbox{$\rm R_{lim}$}}
\def\ms{\mbox{$\rm M_\odot$}}
\def\ds{\mbox{$\rm d_\odot$}}
\def\jj{\mbox{$\rm J$}}
\def\mlow{\mbox{$\rm m_{low}$}}
\def\kms{\mbox{$\rm km\,s^{-1}$}}
\def\mas{\mbox{$\rm mas\,yr^{-1}$}}
\def\fb{\mbox{$\rm f_{bin}$}}
\def\vct{\mbox{$\rm v_{cut}$}}

\begin{document}

\title{Proper motion measurements as indicators of binarity in open clusters}

\author{E. Bica\inst{1} \and C. Bonatto\inst{1}}

\offprints{Ch. Bonatto - charles@if.ufrgs.br}

\institute{Universidade Federal do Rio Grande do Sul, Instituto de F\'\i sica, 
CP\,15051, Porto Alegre 91501-970, RS, Brazil\\
\mail{}
}

\date{Received --; accepted --}

\abstract{We analyze 9 open clusters with ages in the range 70\,Myr to 3.2\,Gyr
using UCAC2 proper motion data and 2MASS photometry, which allows us to 
reach stellar masses down to $\approx0.7\,\ms$. We employ in this work an approach 
in which the background proper motion contribution is statistically subtracted
in order to obtain the cluster's intrinsic proper motion distribution. For each 
cluster we consider the projected velocity distributions in the core and off-core 
regions separately. In the projected velocity distribution of all sample 
clusters we find a well-defined low-velocity peak, as well as an excess in the number of 
stars at larger velocities. The low-velocity peak is accounted for by the random 
motion of the single stars, while the high-velocity excess can be attributed to the 
large velocity changes produced by a significant fraction of unresolved binaries in 
a cluster. We derive kinematic parameters of the single-star distribution, in particular 
the projected velocity dispersion. The relatively large velocity dispersions derived in this
work may reflect the non-virialized state of the clusters. Based on the relative number of 
high-velocity (binary) and single stars, we inferred for the sample clusters unresolved 
binary fractions in the range $15\%\le\fb\le54\%$, for both core and off-core regions. 
Stars with a projected velocity exceeding the maximum reached by the single-star 
distribution are identified in 2MASS $\jj\times\jh$ colour-magnitude diagrams. The 
asymmetry observed in the distribution of these stars around the main sequence is 
consistent with models of main-sequence widening resulting from unresolved binaries 
combined with 2MASS photometric uncertainties. The present results suggest that care
must be taken when applying proper-motion filters to sort out members, especially
binaries in a star cluster. This paper shows that proper motions 
turn out to be a useful tool for identifying high-velocity stars as unresolved binary 
cluster members, and as a consequence, map and quantify the binary component in 
colour-magnitude diagrams.

\keywords{(Galaxy:) open clusters and associations: general} }

\titlerunning{Proper motions in open clusters}

\authorrunning{E. Bica \and C. Bonatto}

\maketitle

\section{Introduction}
\label{intro}

Binaries play a vital r\^ole in the dynamics of star clusters since, as a result of
the combined effect of high star density and low velocity dispersion, the energy
locked up in binary binding energy may be comparable to, or even exceed that of the
whole cluster (Hut \cite{Hut1996}). Consequently, changes in binary properties along 
the normal stellar evolution will affect the dynamical evolution of the cluster as a 
whole, particularly the old, more massive clusters. In this context, the identification 
and quantification of the binary fraction is fundamental for understanding the dynamical 
evolution of star clusters.

Open clusters are known to contain significant fractions of the stellar content in 
the form of multiple systems, binaries in particular (e.g. Montgomery, Marschall \& Janes 
\cite{Montg1993}; von Hippel \& Sarajedini \cite{Hippel1998}). And, as a consequence 
of the dynamical evolution in open clusters, the fraction of multiple systems both 
increases and tends to concentrate in the central regions, thus changing the initial 
spatial distribution of stars (Takahashi \& Portegies Zwart \cite{TakaP2000}). 

Observationally, the main effects of a significant fraction of unresolved binaries in 
the central parts of a star cluster are expected to be: {\em (i)} the production of large 
deviations in the cluster's proper motion distribution with respect to the single-star 
component, since the resulting velocities may be much larger for the binary motions,  
{\em (ii)} the underestimate of the number of low-mass stars with respect to higher-mass 
stars, and {\em (iii)} the widening of the main-sequence (MS) towards higher magnitudes
in a colour-magnitude diagram (CMD). Point {\em (i)} is the fundamental issue addressed 
for the first time in the present paper. As a consequence of {\em (ii)}, the observed 
central luminosity (or mass) function turns out to be flatter than that of the actual, 
resolved stellar distribution. In addition, the resulting mass-function changes will 
depend both on the binary fraction and MS mass range (e.g. Kroupa, Tout \& 
Gilmore \cite{KTG1991} and references therein). 

Rather than consider the proper motion components of individual stars, which might 
exclude unresolved binaries as non-member stars, we apply in this work a statistical 
analysis. Our approach is to compare the proper motion (modulus) distribution of the stars 
in the cluster field with that of the offset field stars. Thus, the presence of unresolved 
binaries should be indicated by significant deviations in the cluster's proper motion 
modulus distribution. 

In the present work we employ J and H 2MASS\footnote{The Two Micron All Sky Survey, All Sky 
data release (Skrutskie et al. \cite{2mass1997}), available at 
{\em http://www.ipac.caltech.edu/2mass/releases/allsky/}} data, which provide photometry 
for stars in a large spatial area in the direction of a cluster, and proper motion data 
from UCAC2\footnote{The Second U.S. Naval Observatory CCD Astrograph Catalog 
(Zacharias et al. \cite{ucac2}), available at {\em http://vizier.u-strasbg.fr/viz-bin/
VizieR?-source=UCAC2}} in the same field. 

This paper is organized as follows. In Sect.~\ref{ttc} we provide general data on 
the target clusters. 2MASS photometry, CMDs and structural parameters for the clusters
are presented in Sect.~\ref{2mass}. The proper motion analysis is presented in
Sect.~\ref{ucac2}. Finally, concluding remarks are given in Sect.~\ref{Conclu}. 

\section{The target clusters}
\label{ttc}

The open clusters selected for the present analysis are M\,26 (NGC\,6694), NGC\,2287,
M\,48 (NGC\,2548), M\,93 (NGC\,2447), NGC\,5822, NGC\,6208, NGC\,3680, IC\,4651 and
M\,67 (NGC\,2682). These clusters span a wide range in age, which is important, since 
the binary fraction in a cluster is expected to increase as a function of dynamical 
evolution (age). General data for the above open clusters are given in Table~\ref{tab1}. 
The data in cols.~2-6 have been obtained from the WEBDA open cluster database (Mermilliod 
\cite{Merm1996})\footnote{{\em http://obswww.unige.ch/webda}}. In cols.~7-11 we list 
parameters derived from the present work using 2MASS data (Sect.~\ref{2mass}).

\begin{table*}
\caption[]{General data on the target clusters.}
\label{tab1}
\renewcommand{\tabcolsep}{1.60mm}
\begin{tabular}{lccccccccccc}
\hline\hline
&\multicolumn{5}{c}{WEBDA}&\multicolumn{5}{c}{Present work}\\
\cline{2-6}\cline{8-12}\\
Cluster&$\alpha(2000)$&$\delta(2000)$&Age&\ebv&\ds&&Age&\ebv&\ds&\rc&\rl \\
&(hms)&($^\circ\arcmin\arcsec$)&(Myr)&&(kpc)&&(Myr)&&(kpc)&(pc)&(pc)\\
(1)&(2)&(3)&(4)&(5)&(6)&&(7)&(8)&(9)&(10)&(11)\\
\hline
M\,26&18:45:18&-09:23:00    &85&0.59&1.60&&$70\pm10$&$0.42\pm0.03$&$1.57\pm0.07$&$0.9\pm0.1$&$7.8\pm0.9$\\
NGC\,2287&06:46:01&-20:45:24&243&0.03&0.69&&$160\pm10$&$0.00$&$0.76\pm0.04$&$1.4\pm0.2$&$8.6\pm0.4$ \\
M\,48&08:13:43&-05:45:00    &360&0.03&0.77&&$360\pm40$&$0.00$&$0.76\pm0.03$&$0.9\pm0.2$&$8.9\pm0.4$ \\
M\,93&07:44:30&-23:51:24    &387&0.05&1.04&&$400\pm50$&$0.00$&$1.05\pm0.04$&$0.6\pm0.1$&$5.5\pm0.6$ \\
NGC\,5822&15:04:21&-54:23:48&662&0.15&0.92&&$1000\pm100$&$0.00$&$0.72\pm0.03$&$0.9\pm0.1$&$8.5\pm0.4$ \\
NGC\,6208&16:49:28&-53:43:42&1170&0.21&0.94&&$1300\pm100$&$0.21\pm0.03$&$1.31\pm0.06$&$0.9\pm0.2$&$6.5\pm0.8$ \\
NGC\,3680&11:25:38&-43:14:36&1200&0.07&0.94&&$1600\pm100$&$0.00$&$1.05\pm0.05$&$0.6\pm0.1$&$7.0\pm0.6$ \\
IC\,4651&17:24:49&-49:56:00 &1140&0.12&0.89&&$1800\pm200$&$0.00$&$0.91\pm0.04$&$0.9\pm0.1$&$5.9\pm0.3$ \\
M\,67 &08:51:18&+11:48:00   &2560&0.06&0.91&&$3200\pm100$&$0.00$&$0.87\pm0.09$&$1.2\pm0.1$&$8.8\pm0.9$ \\
\hline
\end{tabular}
\begin{list}{Table Notes.}
\item The 2MASS colour excess for M\,26 is $\ejh=0.14$, and for NGC\,6208 $\ejh=0.07$. 
Uncertainties in columns (7)-(9) are derived from the isochrone fit.
\end{list}
\end{table*}

In previous studies binaries have already been detected in some of the above clusters.
Stetson (\cite{Stetson1981}) indicated that NGC\,5822 contains a high percentage
of binary stars and several blue stragglers. Levato \& Malaroda (\cite{Levato1979}) 
found that most of the B\,8--A\,0 stars in NGC\,2287 are spectroscopic binaries. 
Nordstr\"om, Andersen \& Andersen (\cite{Nordstrom1997}) identified several single
and binary cluster members in NGC\,3680. With respect to IC\,4651, Meibom, Andersen 
\& Nordstr\"om (\cite{Meibom2002}) found that 37\% of the 19 red giants are spectroscopic 
binaries with periods of up to 5\,000 days, while 52\% of the 67 MS stars are binaries 
with periods less than 1\,000 days. Montgomery, Marschall \& Janes (\cite{Montg1993}) 
found that at least 38\% of the stars in M\,67 are binaries. Mermilliod \& Mayor 
(\cite{MerMay1989}) have found a percentage of 25-33\% of binaries in M\,93. 

\section{The 2MASS photometry}
\label{2mass}

2MASS has proved to be a useful tool to explore various aspects related to open clusters,
e.g. Bica, Bonatto, \& Dutra (\cite{BBD2003}, \cite{BBD2004}).
We derive parameters for the present cluster sample from J and H 2MASS photometry. The VizieR 
tool\footnote{{\em http://vizier.u-strasbg.fr/viz-bin/VizieR?-source=II/246}} was used 
to extract stars in a circular area centered on the coordinates given in Table~\ref{tab1}.
In order to maximize the statistical significance and representativity of background star 
counts, we used stars extracted in an external annulus as offset field.

\subsection{Colour-Magnitude Diagrams}
\label{cmds}

\begin{figure}
\resizebox{\hsize}{!}{\includegraphics{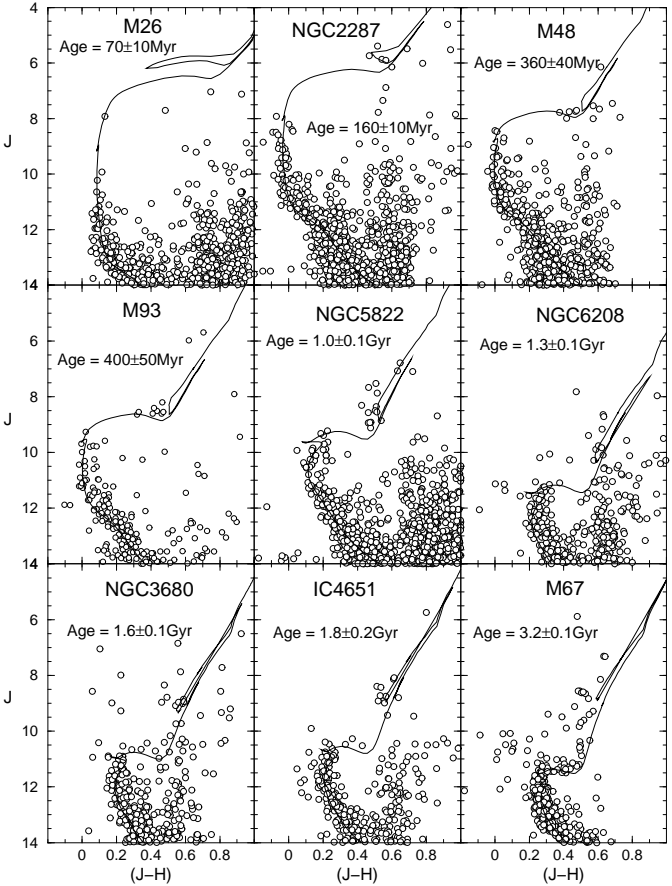}}
\caption[]{$\jj\times\jh$ CMDs for the present open clusters. In each panel, the region
shown corresponds to a radius $r=2\times\rc$. The solid line shows the best-fit Padova
isochrone. The derived age is indicated in each panel.}
\label{fig1}
\end{figure}

In Fig.~\ref{fig1} we present the $\jj\times\jh$ CMDs for the open clusters dealt with 
in this paper. To maximize cluster membership probability, the stars included in the CMDs 
correspond to a region with radius $2\times\rc$ (Sect.~\ref{ClusStr}). Cluster parameters 
were derived by fitting solar metallicity Padova isochrones (Girardi et al. 
\cite{Girardi2002}) computed with the 2MASS J and H filters\footnote{available to download 
at {\em http://pleiadi.pd.astro.it/isoc\_photsys.01/isoc\_photsys.01.html}},
to the observed CMDs. The 2MASS transmission filters produced isochrones very similar to 
the Johnson ones, with differences of at most 0.01 in \jh\ (Bonatto, Bica \& Girardi 
\cite{BBG2004}), see also Grocholski \& Sarajedini (\cite{GS2003}) for an analysis of combined 
optical/near-infrared CMDs. For reddening and absorption transformations we use R$_V$ = 3.2, and the 
relations A$_J = 0.276\times$A$_V$ and $\ejh=0.33\times\ebv$, following Dutra, Santiago 
\& Bica (\cite{DSB2002}) and references therein. The resulting age, colour excess and 
distance to the Sun, are given respectively in cols.~7, 8 and 9 of Table~\ref{tab1}. Except 
for M\,26 and NGC\,6208, no amount of reddening was detected for the clusters.

For most clusters, the optical value for $\ebv$ in Table~\ref{tab1} is larger than
that inferred from the near-infrared. However, we note that $\ebv=3\times\ejh$, so these 
near-infrared bands are in fact less sensitive to reddening determinations. In previous 
studies we already noticed the need for less equivalent reddening in the near-infrared
than what was expected from the optical, e.g. the case of M\,67 (Bonatto \& Bica \cite{BB2003}). 
The excesses \ejh\ are derived from the overall isochrone fit to the CMD sequences. Besides,
for several of the present clusters, deep CMDs are studied. As can be seen in the CMDs of
Fig.~\ref{fig1}, there is not much freedom for variations in \ejh\ measurements. Basically, 
all values of \ebv\ smaller than $0.06$ are null in the near-infrared. In the case of IC\,4651 
(Fig.~\ref{fig1}), the WEBDA value $\ebv=0.12$ would prohibitively shift the isochrone too 
far into the red.

\subsection{Cluster structure}
\label{ClusStr}

Structural parameters for the clusters have been derived by means of the radial distribution
of stars, defined as the projected number of stars per area around the cluster center. 
Before counting stars, we applied colour filters to both cluster and corresponding offset 
field, in order to take account of the contamination of the Galaxy. This procedure has been 
previously applied in the analysis of the open clusters M\,67 (Bonatto \& Bica \cite{BB2003}), 
NGC\,188 (Bonatto, Bica \& Santos Jr. \cite{BBS2004}) and NGC\,3680 and NGC\,2180 (Bonatto, 
Bica \& Pavani \cite{BBP2004}). The radial distribution was determined by counting 
stars inside concentric annuli with a step of 1.0\arcmin\ in radius. The background contribution 
level corresponds to the average number of stars included in the external annulus (offset field).

\begin{figure} 
\resizebox{\hsize}{!}{\includegraphics{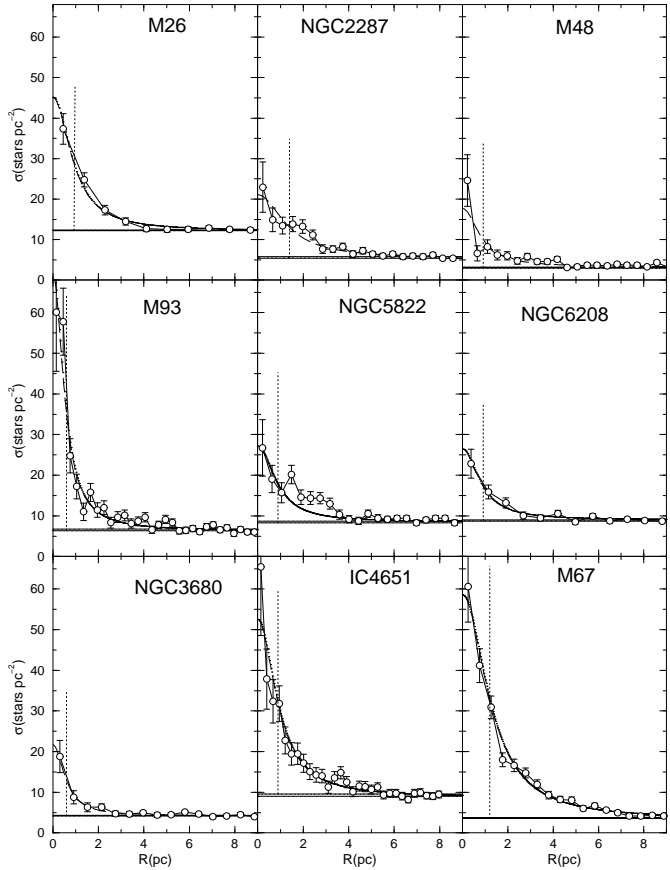}}
\caption[]{Projected radial distributions of stars. For each cluster, the average background 
level is shown as a narrow shaded area; $1\sigma$ Poisson errors are also shown. The dashed 
lines show the two-parameter King model fitted to the radial distributions of stars. In each 
panel, the core radius is indicated as a dotted line. For a better comparison, all panels are 
plotted on the same scale.}
\label{fig2}
\end{figure}

The resulting projected radial density profiles for the present clusters are shown in 
Fig.~\ref{fig2}. For a better comparison between clusters, the radii on the abscissas
have been scaled from arcmin to parsecs using the distances derived in Sect.~\ref{cmds}.
The statistical significance of each profile is reflected by the $1\sigma$ Poisson error 
bars.

First-order structural parameters for each cluster are derived by fitting the two-parameter 
King (\cite{King1966a}) surface density profile to the background-subtracted radial distribution 
of stars. The two-parameter King model essentially describes the central region of 
normal clusters (King \cite{King1966b}; Trager, King \& Djorgovsky \cite{TKD95}). The fits 
were performed using a non-linear least-squares fit routine which uses the errors as 
weights. The best-fit solutions are shown in Fig.~\ref{fig2} as dashed lines, and the resulting 
core radii are given in col.~10 of Table~\ref{tab1}. Considering the profile fluctuations with 
respect to the background level, we can define a limiting radius (\rl) for each cluster. For
regions beyond $\rl$ the null-contrast between cluster and background star 
density would produce exceedingly large Poisson errors and consequently, meaningless 
results. Thus, for practical purposes, we can consider that most of the cluster's stars are 
contained within $\rl$. The limiting radii are given in col.~11 of Table~\ref{tab1}.

\section{Proper motion analysis}
\label{ucac2}

Proper motion components in right ascension and declination for the cluster stars have been 
obtained from UCAC2. Rather than work separately with proper motion components, we employ 
in the present work the projected velocity on the sky $v_p =\sqrt{\mu_\alpha^2\times\cos(\delta)^2+
\mu_\delta^2}$, which corresponds to the modulus of the proper motion components. Although 
affected by the systemic motion, asymmetries in the proper motion distribution of a cluster may 
yield information on the internal kinematics. To be consistent with the 2MASS analysis, we 
extracted from UCAC2 proper motions for each cluster inside the same area as that used to 
extract 2MASS photometry. Since the UCAC2 catalogue also includes the 2MASS photometry, we 
verified that the correspondence of the two catalogues is nearly complete for $\jj\leq 14.5$. 
This is important since for the interpretation of proper motion data in the context 
of determining the physical nature of an object one must include faint stars as well. 

\begin{figure} 
\resizebox{\hsize}{!}{\includegraphics{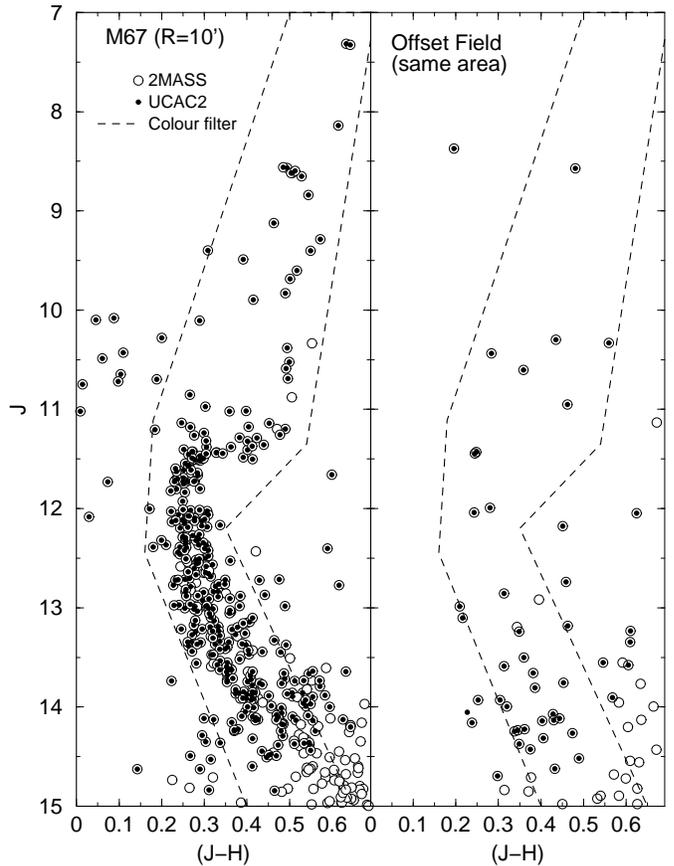}}
\caption[]{Correspondence between the 2MASS photometry and UCAC2 data for M\,67. Both 
catalogues are nearly complete for $\jj<14.5$. The colour filter is shown as dashed
lines. Left panel: central ($R<10\arcmin\approx2.5$\,pc) region; right panel: offset field 
(same area).}
\label{fig3}
\end{figure}

\subsection{M\,67, a test case}
\label{m67}

Because of its relatively high Galactic latitude ($b\approx32^\circ$) the field of M\,67 
(NGC\,2682) presents little background contamination (Fig.~\ref{fig1}). Thus, M\,67 is 
an ideal cluster to examine the probable association of proper motion with the presence 
of binary stars. In this sense, M\,67 is an interesting cluster because it is known to 
contain a significant fraction of unresolved binaries (Montgomery, Marschall \& Janes 
\cite{Montg1993}). In Fig.~\ref{fig3} we show the $\jj\times\jh$ CMDs of the central 
region of M\,67 (left panel) and the corresponding (same area) offset field (right panel). 
Stars with 2MASS detections are indicated by open circles, while those with proper motion 
determined by UCAC2 are indicated by dots. For stars with $\jj\leq 14.5$ both catalogues 
are essentially complete. We also show in both panels the colour filter used to isolate
most M\,67 stars from the background contamination. Probable M\,67 blue stragglers (at
$10\leq\jj\leq11$ and $0.0\leq\jh\leq0.3$) have also been excluded from the analysis.
The relatively large number of stars in M\,67 also allows us to study spatial variations 
in the proper motion distribution. Thus, the present analysis will be applied to both core 
and off-core regions.

\begin{figure} 
\resizebox{\hsize}{!}{\includegraphics{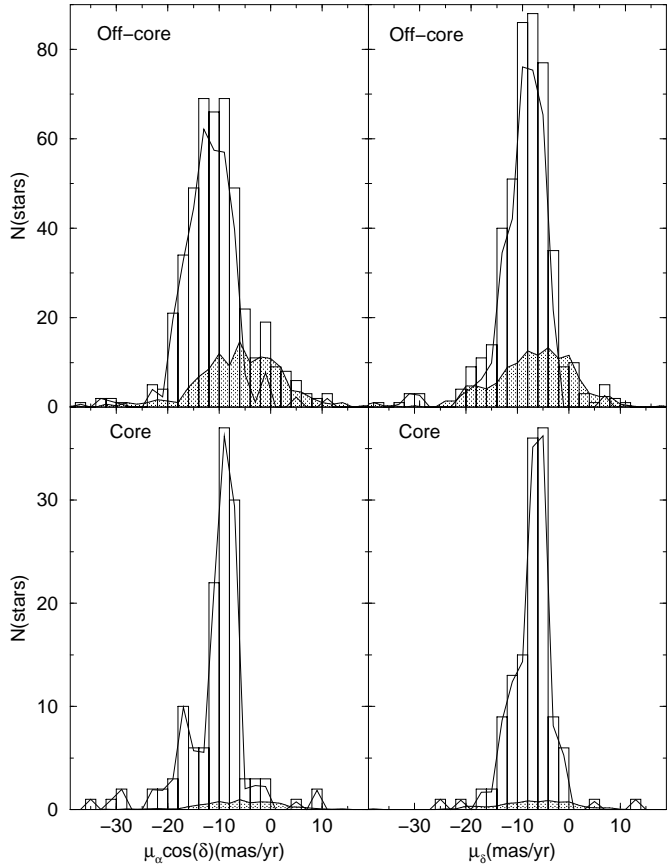}}
\caption[]{Proper motion components in the core (bottom panels) and off-core (top
panels) regions of M\,67. Histograms in the left panels show the number of stars per 
velocity bin for the $\mu_\alpha$ component; $\mu_\delta$ components are in the right panels.
The shaded areas correspond to the offset field (same area) contribution. The solid line
represents the intrinsic proper motion distribution in each region.}
\label{fig4}
\end{figure}

Although small, the contamination by Galactic stars of the CMD of M\,67 must be taken into 
account to isolate the intrinsic proper motion distribution in this cluster.
We study the properties of the proper motions in M\,67 according to the following
procedure. First we apply the colour filter shown in Fig.~\ref{fig3} to both cluster and
offset field stars. The filtering process takes into account most of the background (mostly
Galaxy disk stars) contamination, leaving only a residual contamination which will be dealt with
by the offset field. Next we build histograms with the number of stars in velocity bins of 
$2\,\mas$ width, both for the cluster and the offset field. This procedure is applied to both 
core ($0.0\le R({\rm pc})\le 1.2$) and off-core ($1.2\le R({\rm pc})\le 8.8$) regions. Then, the offset 
field histogram is scaled to match the projected area of both regions. Finally, the subtraction 
of the offset field histogram from those of the core and off-core regions produces the intrinsic 
proper motion distribution in M\,67. This procedure is illustrated in Fig.~\ref{fig4}
for the proper motion components in right ascension ($\mu_\alpha$) and declination 
($\mu_\delta$). As expected for a self-gravitating system, the proper motion distributions
in the core and off-core regions of M\,67 present systematic deviations with respect to the 
offset field distribution. 

\begin{figure} 
\resizebox{\hsize}{!}{\includegraphics{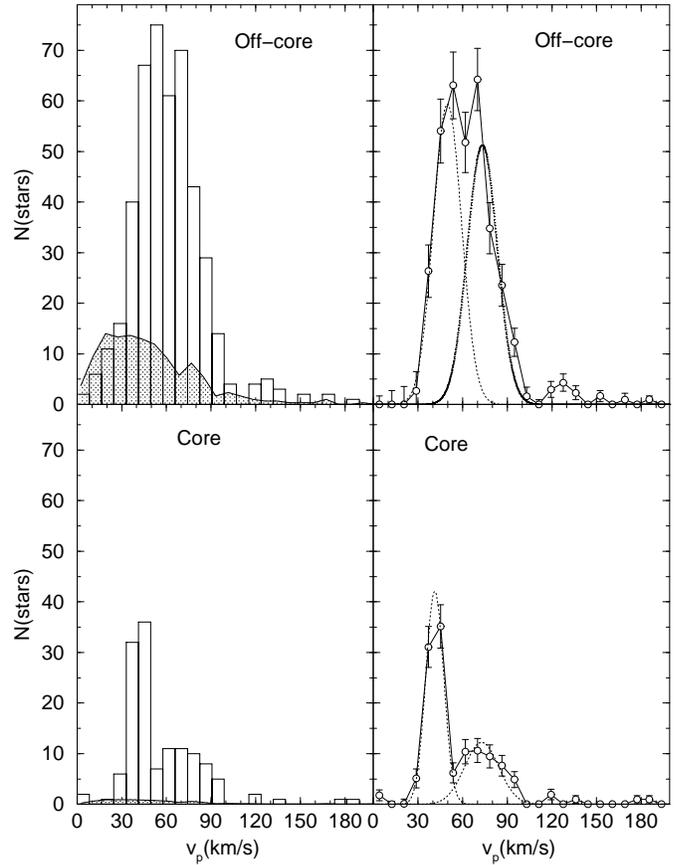}}
\caption[]{Distribution of projected velocity of stars in M\,67. Histograms in the
left panels show the number of stars per velocity bin in the core (bottom
panel) and off-core regions (top panel). The shaded area corresponds to the offset
field (same area) contribution. Right panels: projected velocity distribution.
Dotted lines: Gaussian fits to the distributions. Note an additional high-velocity
peak centered at $v_p\approx130\,\kms$ in the off-core profile of M\,67.}
\label{fig5}
\end{figure}

Significant deviations in proper motion with respect to a Gaussian distribution are seen 
in both $\mu_\alpha$ and $\mu_\delta$ components. Considering the offset field contribution,
the excesses may be accounted for by a component of high-velocity stars inside M\,67. In 
order to characterize these deviations in terms of high-velocity stars, we hereafter use the 
projected velocity $v_p$. The units $\mas$ have been transformed to \kms\ using the 
distances listed in col.~9 of Table~\ref{tab1}. The above procedure is illustrated in 
Fig.~\ref{fig5}, in which the left panels show the projected velocity histograms both for 
the cluster regions and offset-field, while the right panels display the intrinsic, offset 
field-subtracted, projected velocity distribution in M\,67. The statistical significance of 
this procedure can be estimated from the resulting Poisson error bars. 

The projected velocity distributions both in the core and off-core regions of M\,67 present
conspicuous low-velocity and high-velocity peaks. The low-velocity peaks can be accounted
for by the random motion of the single stars, superimposed on the cluster's systemic 
motion. On the other hand, the high-velocity peaks, which considering the error bars represent 
real excesses, may be produced by unresolved binary systems, in which the presence of a
secondary changes appreciably the velocity of the primary star. Kinematical parameters for the 
cluster are derived by fitting Gaussian functions $N\propto\exp{-\frac{(v-v_0)^2}{2\sigma_v^2}}$ 
to the $v_p$ distributions, where $v_0$ is the peak average velocity and $\sigma_v$ is the 
projected velocity dispersion. We found for the core region a low-velocity peak centered at 
$v_0=41.3\pm0.7\,\kms$, with a velocity dispersion $\sigma_v=6.0\pm0.6\,\kms$. The high-velocity 
peak, with smaller amplitude, is centered at $v_0=72.8\pm2.3\,\kms$, with $\sigma_v=11.5\pm1.5\,\kms$. 
A similar situation is observed in the off-core region, with the low-velocity peak at 
$v_0=49.8\pm1.8\,\kms$, with $\sigma_v\approx7.5\pm2.3\,\kms$, and the high-velocity one at 
$v_0\approx73.0\pm2.2\,\kms$, with $\sigma_v\approx10.1\pm0.7\,\kms$. The above fits are shown 
in the right panels of Fig.~\ref{fig5} as dotted lines. Within the uncertainties, $\sigma_v$ for 
the low-velocity peak is the same, both in the core and off-core regions. The same conclusion
applies to the high-velocity peaks in both regions. 

\begin{figure} 
\resizebox{\hsize}{!}{\includegraphics{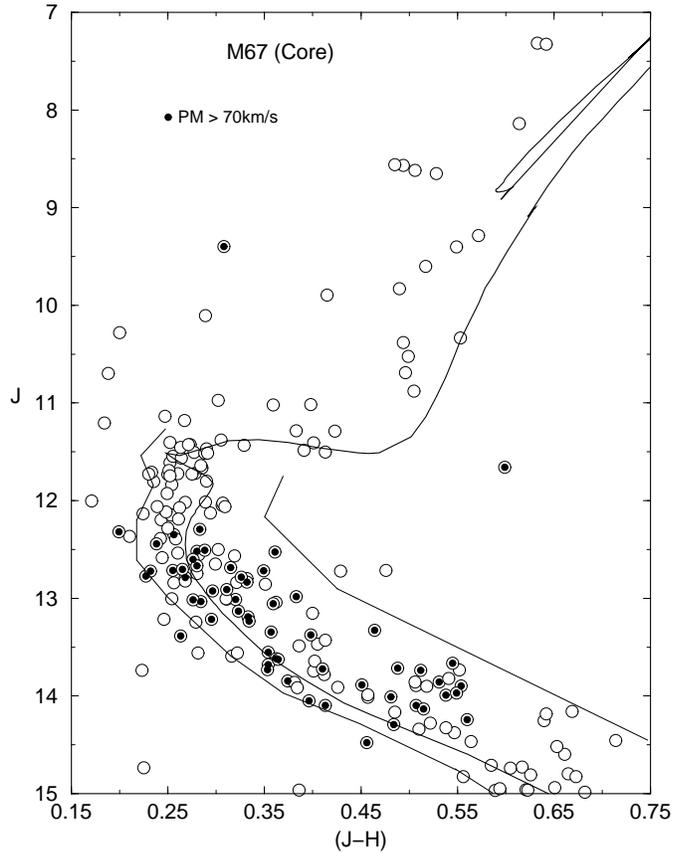}}
\caption[]{Stars with projected velocity larger than $70\,\kms$ (filled circles) are 
superimposed on the CMD of M\,67. The thin solid lines mark the borders of the CMD  
region occupied by unresolved binaries combined to the 2MASS photometric uncertainties. 
The $3.2\,$Gyr isochrone is also shown for clarity.}
\label{fig6}
\end{figure}

Considering the $v_p$ distribution in the core region, stars with velocities larger 
than $70\,\kms$ can be considered to have deviant proper motions with respect to the single-star 
distribution. We identify and plot these deviant stars in the CMD of the core region in 
Fig.~\ref{fig6}, in which the $3.2\,$Gyr isochrone is also plotted for clarity. The
high-velocity stars are distributed along the MS with an asymmetry towards the right side of the 
isochrone. They occupy the locus encompassed by the colour-magnitude limits expected for binary 
stars combined with the 2MASS photometric uncertainties (thin-solid lines in Fig.~\ref{fig6}).

To simulate the effects of unresolved binaries on the 2MASS CMD, we first generate a population 
of single stars according to a standard ($\chi=1.35$) Salpeter MF distribution. To be consistent 
with the 2MASS CMD of M\,67 we used single stars with masses lower than the respective turnoff. 
According to a pre-defined fraction of CMD binaries, we randomly select stars from the 
original distribution and build the specified number of binaries. No mass or luminosity bias is 
introduced in this process (Kroupa \cite{Kroupa2002}; Mazeh et al. \cite{Mazeh2003}) and, consequently 
the probability of a star being selected as a binary member depends only on the number frequency 
of its mass. In order to obtain the maximum spread in \jh\ colour spanned by the binary 
distribution we use a binary fraction of 100\%, which means that all objects in the CMD 
are unresolved binary systems. We also include in the simulation the effects of the 2MASS 
photometric uncertainties in J and H. In 2MASS photometry, the average photometric uncertainty 
($\epsilon$) for a given magnitude can be obtained following $\epsilon_J=0.023+4.78
\times10^{-9}e^{(J/0.976)}$ and $\epsilon_H=0.025+1.92\times10^{-8}e^{(H/0.998)}$ (Bonatto, Bica 
\& Santos Jr. \cite{BBS2004}). For each J and H magnitude, the photometric error is obtained by 
randomly selecting a value in the interval ($-\epsilon_{J,H}\leq\epsilon\leq+\epsilon_{J,H}$), 
assuming that $\epsilon$ follows a normal distribution curve. Finally, for each single-mass star 
and binary in the CMD, uncertainties are assigned to the J and H magnitudes as described above.
The resulting maximum colour spreads are indicated in Fig.~\ref{fig6}. The high-velocity
stars in Fig.~\ref{fig6} do not reach the red border of the simulation. Consequently, the
binary fraction must be lower than 100\%.

Under the assumption that the low-velocity peak essentially represents the velocity distribution
of the single stars, and that the high-velocity excess is caused by unresolved binaries,
the binary fraction (\fb) in a cluster can be estimated by dividing the number of high-velocity
stars by the total number of stars. Thus, in the core region of M\,67, we find $\fb=39\pm16\%$,
which agrees with the fraction derived by Montgomery, Marschall \& Janes (\cite{Montg1993}). 
In the off-core region we find $\fb=45\pm12\%$ which, within the uncertainties, agrees with the
binary fraction in the core.

\subsection{The remaining sample}
\label{other}

\begin{figure} 
\resizebox{\hsize}{!}{\includegraphics{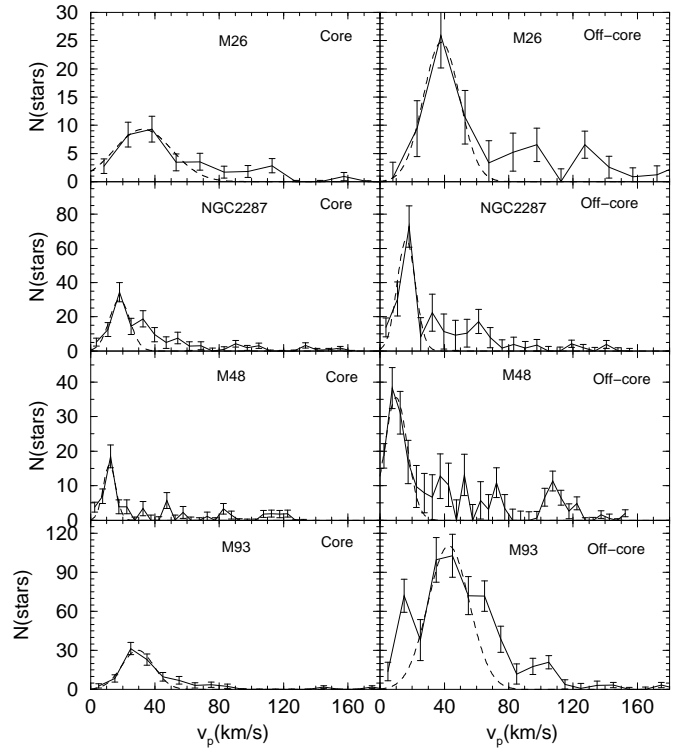}}
\caption[]{Projected velocity distributions in the core (left panels) and
off-core (right panels) regions. The dashed curves represent a fit to the
single-star distributions, according to the parameters in Table~\ref{tab2}.}
\label{fig7}
\end{figure}

\begin{figure} 
\resizebox{\hsize}{!}{\includegraphics{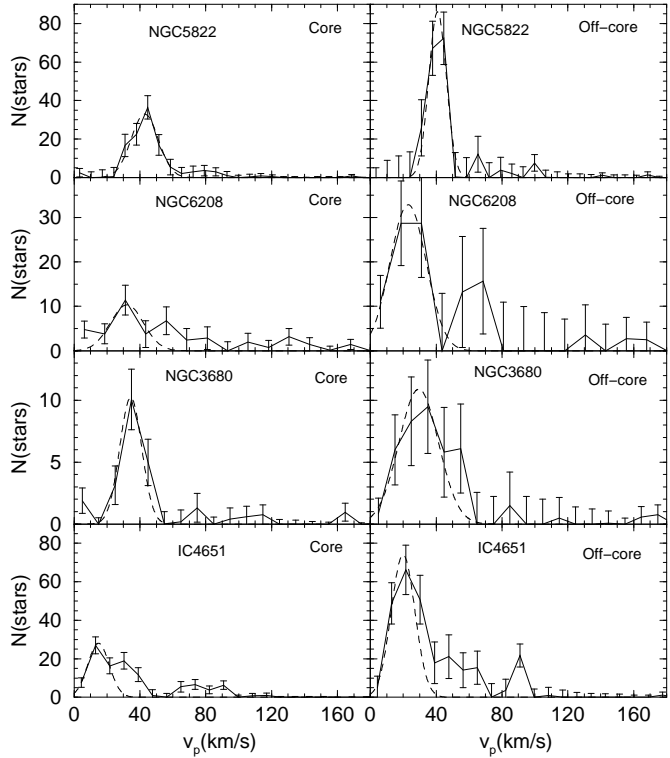}}
\caption[]{Same as Fig.~\ref{fig7} for the remaining clusters.}
\label{fig8}
\end{figure}

\begin{figure} 
\resizebox{\hsize}{!}{\includegraphics{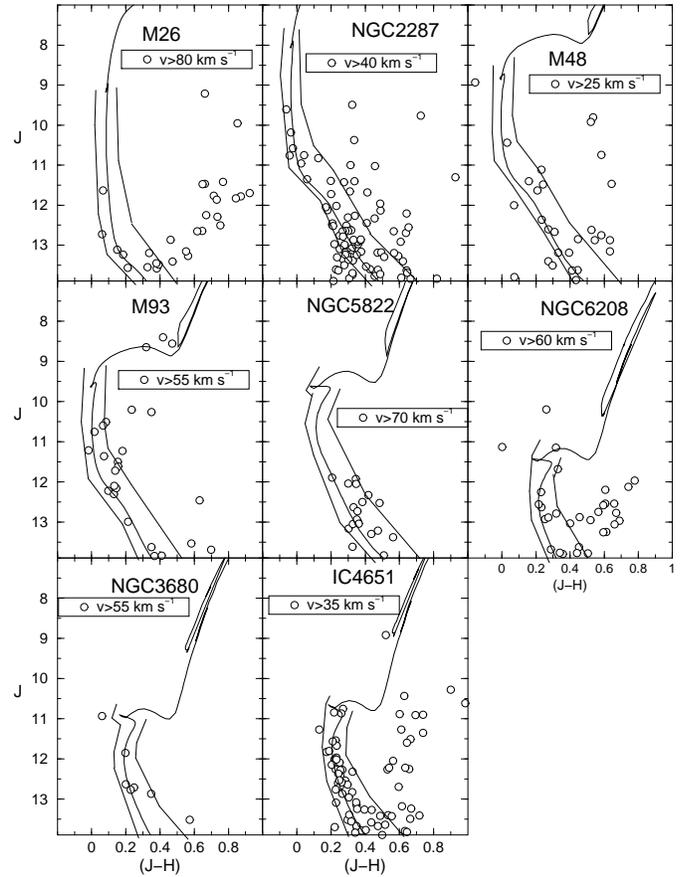}}
\caption[]{Distribution of the high-velocity stars in the core region of
the remaining clusters. The minimum value for a high-velocity star
(\vct) is indicated in each panel. For ease of visualization, only the 
isochrone and the borders of the binary+photometric error simulation are shown.}
\label{fig9}
\end{figure}

The remaining clusters have been analyzed in a similar way as M\,67, and the corresponding 
projected velocity distributions for the core and off-core regions are shown in Figs.~\ref{fig7} 
and \ref{fig8}. Clearly, the low-velocity peak is present in both core and off-core regions in 
all clusters, with varying central (systemic) velocity and velocity dispersion. This is 
reflected in the parameters of the Gaussian fits to the low-velocity peaks, which are summarized 
in Table~\ref{tab2}. We include in Table~\ref{tab2} the maximum velocity reached in the 
single-star distribution, \vct. The high-velocity excess is conspicuously present 
in most clusters as well, the exceptions being NGC\,3680 and NGC\,5822 (core and off-core regions), and
M\,48 and M\,93 (core region). The large error bars present in the profiles of the latter clusters, 
which reflect a relatively small number of stars, make the excess marginal. Gaussian fits have 
not been applied to the high-velocity excess in the remaining clusters because of the less well
defined profiles. Besides, as a consequence of the large error bars the resulting binary fractions 
contain significant uncertainties. Within the uncertainties, the average binary fraction in the 
core, $<\fb>=35\pm15\%$, turns out to be the same as in the off-core region, $<\fb>=36\pm11\%$ 
(standard deviation from the average). The same is valid for the average velocity dispersions 
in the core, $<\sigma_v>=8\pm4\,\kms$, and off-core regions, $<\sigma_v>=9\pm2\,\kms$. The 
high-velocity stars (according to col.~10 of Table~\ref{tab2}) that are present in the core 
region of each of the remaining clusters are identified and plotted on the corresponding CMDs 
in Fig.~\ref{fig9}. For each cluster we simulated the combined effect of binaries and 
photometric errors, following the approach applied to M\,67 (Sect.~\ref{m67}). The resulting 
limiting lines are shown in the figure. In most clusters the asymmetry in colour presented by 
the high-velocity stars with respect to the corresponding isochrone is similar to that 
displayed by the core of M\,67 (Fig.~\ref{fig6}), bearing in mind that M\,67 is more populous. 
The observed asymmetries in the CMD distributions support the scenario of unresolved 
binaries, as further indicated by the present proper motion study.

In the CMD of some clusters, a nearly vertical red sequence shows up (e.g. IC\,4651) which we 
suspect to be solar neighbourhood high-velocity low-mass stars. From their \jh\ colours, we infer 
them to be K and M dwarfs. Similar to the case of M\,67, most of the high-velocity stars do fall 
within the red border of the binary simulation, except in M\,26 and NGC\,6208 which are much 
affected by red stars.

\begin{table*}
\caption[]{Properties of the single-star distribution and binaries.}
\label{tab2}
\renewcommand{\tabcolsep}{2.6mm}
\begin{tabular}{lccccccccccc}
\hline\hline
&&\multicolumn{3}{c}{Off-core}&&\multicolumn{5}{c}{Core}\\
\cline{3-6}\cline{8-11}\\
Cluster&\mlow&$v_0$&$\sigma_v$&\fb&\vct&&$v_0$&$\sigma_v$&\fb&\vct\\
 &(\ms)&(\kms)&(\kms)&(\%)&(\kms)&&(\kms)&(\kms)&(\%)&(\kms)\\
    (1)&(2)&(3)&(4)&(5)&(6)&&(7)&(8)&(9)&(10)\\
\hline
M\,26    &1.1&$38.2\pm1.0$&$11.3\pm0.8$&$39\pm21$&70&&$32.6\pm3.6$&$18.0\pm3.8$&$18\pm12$&80\\
NGC\,2287&0.6&$16.3\pm2.5$&$ 5.7\pm2.2$&$43\pm59$&35&&$18.0\pm1.8$&$6.4\pm1.7$&$48\pm45$&40\\
M\,48    &0.6&$10.0\pm1.5$&$7.0\pm1.8$&$47\pm21$&35&&$11.7\pm0.8$&$3.9\pm0.8$&$48\pm23$& 25\\
M\,93    &0.8&$42.7\pm1.2$&$12.8\pm1.6$&$38\pm17$&85&&$29.3\pm1.3$&$9.9\pm1.2$&$21\pm9$&55\\
NGC\,5822&0.6&$41.1\pm1.3$&$ 8.2\pm1.1$&$16\pm10$&60&&$43.2\pm1.4$&$5.3\pm1.1$&$16\pm8$&70\\
NGC\,6208&0.8&$23.0\pm2.3$&$11.0\pm2.0$&$34\pm22$&60&&$32.1\pm1.9$&$10.1\pm1.3$&$54\pm30$&60\\
NGC\,3680&0.8&$29.4\pm1.8$&$12.0\pm1.5$&$20\pm11$&65&&$34.4\pm0.4$&$6.7\pm0.3$&$25\pm5$&55\\
IC\,4651 &0.7&$19.9\pm2.3$&$ 7.2\pm2.1$&$46\pm50$&45&&$14.8\pm0.5$&$6.5\pm0.5$&$50\pm11$&35\\
M\,67    &0.7&$49.8\pm1.8$&$7.5\pm2.3$&$45\pm12$&90&&$41.3\pm0.7$&$6.0\pm0.6$&$39\pm16$&70\\
\hline
\end{tabular}
\begin{list}{Table Notes.}
\item $\mlow$ is the stellar low-mass end reached in this work; $v_0$ and $\sigma_v$ have been 
obtained from the Gaussian fit to the low-velocity distribution; \fb\ is the binary fraction; 
\vct\ is the maximum velocity reached by the single-star distribution.
\end{list}
\end{table*}

According to the virial theorem, the dynamical mass for a gravitationally bound, spherically 
symmetric and isotropic cluster is given by $M=3\sigma^2R/G$, where $\sigma$ is the velocity
dispersion and R is the effective gravitational radius. As a first-order approximation, we 
assume $R=\rc$ and $\sigma=\sigma_v(core)$. For the range of core radii (Table~\ref{tab1}) and 
velocity dispersions (Table~\ref{tab2}) found in the present cluster sample the implied dynamical 
masses would be in the range $(1-4)\times10^4\,\ms$. For M\,26, the youngest cluster in our 
sample with the largest $\sigma_v$, a mass would result of $\sim2\times10^5\,\ms$. These mass 
values turn out to be exceedingly high for open clusters, since the old ($\approx7\,$Gyr) and 
very populous open cluster NGC\,188 has an estimated mass of $\approx4\times10^3\,\ms$, as 
determined from different methods (Bonatto, Bica, \& Santos Jr. \cite{BBS2004}, and references 
therein). The present clusters are clearly not as populous as NGC\,188. Consequently, the 
discrepancy in the mass determinations certainly reflects the non-virialization of the clusters.

The present sample contains one cluster (NGC\,2287) in common with the clusters studied in 
Sagar \& Bhatt (\cite{SagBhatt89}). Their analysis was based on radial velocity and proper 
motion measurements of individual stars, restricted to the brighter, higher-mass cluster 
members. In contrast, the present 2MASS photometry depth allowed us to reach low-mass 
stars as well, as can be seen in col.~2 of Table~\ref{tab2}.

Under the first-order assumption of energy equipartition and different mass groups within a 
cluster, the velocity dispersion would follow the relation $\sigma\sim1/\sqrt{\overline{m}}$, 
where $\overline{m}$ is the average mass of each stellar group. Thus, Sagar \& Bhatt's
(\cite{SagBhatt89}) results should favour lower velocity dispersions than the present
work. Indeed, the velocity dispersions derived in the present work for the core and off-core 
regions of NGC\,2287 are larger than those found by Sagar \& Bhatt (\cite{SagBhatt89}) in 
similar regions, by a factor of $\approx6$ in the core and $\approx3.4$ in the off-core region. 
The significant differences in velocity dispersion in both studies may be accounted for by the 
larger number of stars, $\approx357$ -- including stars as faint as $\jj\approx15$ and mass 
$m\geq0.6\,\ms$ -- in the present work, as compared to $\approx65$ ($m\geq1.7\,\ms$) stars
in Sagar \& Bhatt's (\cite{SagBhatt89}) sample. The large number of stars allowed us to fit 
the appropriate Gaussian to the background-subtracted projected velocity distribution of 
NGC\,2287 (Fig.~\ref{fig7}). Thus, for such a relatively young cluster as NGC\,2287, in which the 
low-mass stars are expected to be far from virialized, the velocity dispersion derived from 
a large mass range should be larger than that obtained from higher-mass stars. In addition,
Sagar \& Bhatt (\cite{SagBhatt89}) derive, for the very young, non-virialized open clusters 
NGC\,2669 and NGC\,4755, velocity dispersions of the order of 10\,\kms. Interestingly, 
the velocity dispersion in the core of M\,26, the youngest cluster in the present sample 
($\approx70\,$Myr), turns out to be the largest, which agrees with strong non-virialization.

\section{Concluding remarks}
\label{Conclu}

In this paper we have analyzed 9 Galactic open clusters, spanning a range of ages 
from 70\,Myr to 3.2\,Gyr, in terms of UCAC2 proper motion data and 2MASS J and H 
photometry. Considering the depth of the 2MASS photometry for the stars with UCAC2
data, the stellar low-mass end in the present analysis reached down to $\approx0.7\,\ms$. 
Rather than considering individual stars, we applied in this work a statistical approach 
in which the cluster's intrinsic proper motion distribution was obtained taking into 
account the background proper motion contribution. Our goals were to infer the intrinsic 
binary fraction in the clusters and derive kinematic parameters of the single stars, the 
line-of-sight velocity dispersion in particular. The large velocity dispersions derived 
in the present work may reflect the non-virialization of the clusters, particularly with 
respect to the low-mass stellar component.
For each cluster we considered separately the projected velocity 
$\left(v_p =\sqrt{\mu_\alpha^2\times\cos(\delta)^2+\mu_\delta^2}\right)$ distributions 
in the core and off-core regions. We found in all clusters $v_p$ distributions which 
are characterized by a conspicuous low-velocity peak, as well as an excess in the number 
of stars for larger velocities, in varying proportions for different clusters. The 
low-velocity peak can be accounted for by the random motion of the single stars 
superimposed on the cluster's systemic motion. The high-velocity excess is attributed 
to the significant velocity changes produced by the presence of unresolved binaries in 
the clusters. The binary fraction is inferred by taking into account the relative number 
of high-velocity and single stars. We derived a fraction of unresolved binaries in the 
range $15\%\le\fb\le54\%$, for both core and off-core regions in the sample clusters.
We showed that the asymmetry observed in the distribution of high-velocity stars in 
2MASS $\jj\times\jh$ CMDs is consistent with models of MS widening resulting from 
unresolved binaries associated to 2MASS photometric uncertainties. 

A possible application of the present method could be in a discussion of the low binary  
fraction found in the globular cluster M\,4 (1-2\%), a surprising result as pointed out 
by Richer et al. (\cite{Richer2004}), when compared to previous estimates for globular 
clusters, which amount to $\leq10\,\%$ (Richer et al. \cite{Richer1997}). Richer et al. 
(\cite{Richer2004}) applied to M\,4 a proper-motion filtering using multi-epoch HST images 
over 6 years. Possibly in this method a significant fraction of binaries was overlooked 
because large-proper-motion stars were discarded.

We conclude that proper motions are a useful tool for identifying high-velocity stars as 
unresolved binary cluster members, and as a consequence, mapping and quantifying the binary 
component in CMDs.

\begin{acknowledgements}
This publication makes use of data products from the Two Micron All Sky Survey, which 
is a joint project of the University of Massachusetts and the Infrared Processing and 
Analysis Center/California Institute of Technology, funded by the National Aeronautics 
and Space Administration and the National Science Foundation. We also made use of proper 
motion data from UCAC2 (The Second U.S. Naval Observatory CCD Astrograph Catalog) as
well as from the WEBDA open cluster database. We acknowledge support from the Brazilian 
Institution CNPq.
\end{acknowledgements}

%

\end{document}